\documentclass{statsoc}

\usepackage{amsmath}
\usepackage{amssymb}
\usepackage{graphicx}
\usepackage{enumerate}
\usepackage{multirow}
\PassOptionsToPackage{authoryear,sort}{natbib}
\usepackage{shortcuts}
\usepackage{booktabs}
\usepackage[capitalise]{cleveref}
\usepackage{suffix}
\usepackage{enumitem}
\usepackage{bm}
\newcommand{\subparagraph}{}
\usepackage[compact]{titlesec}
\newcommand{\ts}{\textstyle}
\newcommand{\bv}{{\bm v}}
\newcommand{\bX}{{\bm X}}
\newcommand{\bY}{{\bm Y}}
\newcommand{\bW}{{\bm W}}
\newcommand{\bE}{{\bm \epsilon}}
\newcommand{\bmu}{{\bm \mu}}
\newcommand{\bd}{C}
\newcommand{\cv}{\Omega}
\newcommand{\W}{\mathcal W}
\newcommand{\Q}{\mathcal Q}
\newcommand{\K}{\mathcal K}
\newcommand{\F}{\mathcal F}
\newcommand{\X}{\mathcal X}
\newcommand{\M}{\mathcal M}
\newcommand{\B}{\mathcal B}
\newcommand{\Des}{\mathcal S}
\newcommand{\sumi}{\sum_{i=1}^n}

\newcommand{\obs}{{\text{obs}}}
\newcommand{\var}{\op{var}}
\newcommand{\varb}[1]{\var\bracks{#1}}

\title[On the Optimality of Randomization in Experimental Design]{On the Optimality of Randomization in Experimental Design:\\How to Randomize for Minimax Variance and Design-Based Inference}

\author[Nathan Kallus]{Nathan Kallus}
\address{
Cornell University,
New York, NY,
USA.
}
\email{kallus@cornell.edu}

\usepackage{geometry}
\begin{document}
\pdfpageheight 29.7cm
\pdfpagewidth 21.0cm
\begin{abstract}
I study the minimax-optimal design for a two-arm controlled experiment where conditional mean outcomes may vary in a given set.
When this set is permutation symmetric, the optimal design is complete randomization, and using a single partition (\ie, the design that only randomizes the treatment labels for each side of the partition) has minimax risk larger by a factor of $n-1$. More generally, the optimal design is shown to be the mixed-strategy optimal design (MSOD) of \citet{kallus2018optimal}. Notably, even when the set of conditional mean outcomes has structure (\ie, is not permutation symmetric), being minimax-optimal for variance still requires randomization \emph{beyond} a single partition. Nonetheless, since this targets precision, it may still not ensure sufficient uniformity in randomization to enable randomization (\ie, design-based) inference by Fisher's exact test to appropriately detect violations of null. I therefore propose the inference-constrained MSOD, which is minimax-optimal among all designs subject to such uniformity constraints. On the way, I discuss \citet{johanssoncomment} who recently compared rerandomization of \citet{morgan2012rerandomization} and the pure-strategy optimal design (PSOD) of \citet{kallus2018optimal}. I point out some errors therein and set straight that randomization \emph{is} minimax-optimal and that the ``no free lunch'' theorem and example in \citet{kallus2018optimal} are correct.
\end{abstract}

\section{Introduction}

Controlled experimentation is the gold standard for causal inference due to the simple fact that it can make experimental groups statistically equivalent in all ways but for the treatment assignment, so that any observed differences can only be attributed to a causal effect or to noise. The less noise, the more certain we are that differences attest to a true causal effect. Therefore, in the pursuit of scientific discoveries, it behooves us to minimize noise. One way to do so is to increase the number of units, but this is often expensive and so an economical researcher should seek to eliminate as much noise as possible on a given budget of units.

Noise in experiments arises both due to the randomness in sampling units from a population and due to the randomness of treatment assignment. The latter, of course, is completely within the control of the researcher and is often called the design. Therefore, conditional on the sample (or, similarly, if we treat sampling as non-random), we should seek the design with minimal noise. 

A common design is complete randomization, where a random subset of fixed size is chosen for treatment. However, such a design may well result in an assignment that appears ``imbalanced.'' An oft-quoted criticism by Student (W. Gosset) is that it ``would be pedantic to continue with an arrangement of [field] plots known beforehand to likely lead to a misleading conclusion,'' referring to completely randomized experiments in agriculture \citep{gosset}. Both the judgment of ``imbalance'' and the supposed foreknowledge of misleading conclusions, however, must depend on some understanding of how post-treatment outcomes depend on pre-treatment variables. Student, for example, mentions experimental group disparities in average ``fertility slopes.''

Recently, \citet{kallus2018optimal} developed a systematic framework of how such a priori understanding on the structure of this relationship translates to optimal design, using the lens of a priori (meaning, before randomization and treatment) balance in pre-treatment variables. 
The framework is phrased as a zero-sum game against Nature, where the experimenter seeks to eliminate noise and Nature interferes adversarially but is constrained by the assumed structure.
In the absence of structure, it is in fact \emph{impossible} to improve upon complete randomization, referred to by \citet{kallus2018optimal} as ``no free lunch.''
When one assumes structure in the form of restrictions on the conditional expectation function (CEF) of post-treatment outcomes given pre-treatment variables, other (still randomized) designs become optimal.
In particular, \citet{kallus2018optimal} formalizes ``noise'' as 
post-treatment estimation \emph{variance} so these designs
minimize worst-case variance, or are minimax with respect to expected squared error.

In this paper, I re-emphasize how minimax-variance optimal design \emph{does not} imply no randomization.
To elucidate this, I re-introduce the framework of \citet{kallus2018optimal} in a simple, instructive manner that more clearly highlights its minimax structure.
In particular, even when one assumes CEF structure, I demonstrate that \emph{optimal} designs are still \emph{randomized}, even beyond the random flipping of ``treatment'' and ``control'' labels on the experimental groups. I discuss the optimality of such randomization and in what limited cases is randomizing only the treatment label on a single partition of units optimal.
Furthermore, I show how one can {correctly} trade-off balance and additional randomization by relaxing the assumed CEF structure.
I additionally revisit the question of randomization (\ie, design-based) inference and present a new constrained-optimization formulation to find the minimax-optimal design subject to a uniformity constraint to enable randomization testing at a given significance level.

On the way, I also discuss to \citet{johanssoncomment} who recently compared rerandomization \break\citep{morgan2012rerandomization} and the pure-strategy optimal design (PSOD) of \citet{kallus2018optimal}, which is a heuristic offered for the mixed-strategy optimal design (MSOD), which is the minimax-optimal design.
I thank and congratulate \citet{johanssoncomment} for a thought-provoking paper.
I use the opportunity in this paper to set straight a few errors I found in it: randomization beyond just treatment blinding \emph{is} in fact minimax variance optimal; \citet{kallus2018optimal} proposes the MSOD as the minimax-optimal design, which \emph{does} randomize beyond two symmetric assignments; designing for optimal precision subject to enabling randomization inference does not require uniform randomization over a restricted set; optimal schemes as I propose herein exist; and finally Theorem 1 (``no free lunch'') and Example 1 of \citet{kallus2018optimal} are correct (and are misquoted by \citealp{johanssoncomment}) showing that, \emph{in the worst-case}, randomizing between two symmetric assignments, and in particular optimizing the Mahalanobis distance between experimental group means, can increase variance by a factor of $n-1$ relative to complete randomization.

\section{The Framework of \citet{kallus2018optimal} Redux and Refined}

We briefly review the framework of \citet{kallus2018optimal}, presenting it anew more clearly as minimax over parameters and focusing on the case of two treatment arms. First, we set up the problem. Our sample consists of $n$ units with (observed) pretreatment variables $X_i\in\X$ and (unobserved) potential outcomes $Y_i(1),Y_i(-1)\in\Rl$, for $i=1,\dots,n$. Each unit $(X_i,Y_i(1),Y_i(-1))$ is assumed independent of others (but not necessarily identically distributed). Define $\mu_i=\Eb{Y_i(1)+Y(-1)\mid X_i}$ and $\epsilon_i={Y_i(1)+Y(-1)}-\mu_i$.

We are interested in estimating and making inferences on the sample average treatment effect (SATE): $\tau=\frac1n\sumi(Y_i(1)-Y_i(-1))$.
Toward that end, we can choose treatment assignments $\bW=(W_1,\dots,W_n)\in\{-1,1\}^n$ and get to observe $Y^\obs_i=Y_i(W_i)$.
We refer to treatment $1$ as ``treatment'' and $-1$ as ``control.''
(Note also our convention of bold type for tuples.) 
For simplicity, suppose $n$ is even and that $\ip{\bW}{\bm 1}=0$, where $\bm1$ is the vector of ones.
We focus on the SATE estimator $\hat\tau=\frac2n\ip{\bW}{\bY^\obs}$.
Since outcomes are not observed before treatment, $\bW$ must be independent of $(\bY(-1),\bY(1))$ given $\bX$.

A \emph{design} is a distribution $\sigma$ over $\bW$, which specifies how we choose the treatment assignment \emph{conditional} on $\bX$; we treat $\sigma$ as a random variable measurable with respect to $\bX$. We require that every assignment $\bW$ supported by the design has $\ip{\bW}{\bm 1}=0$, and that $\bW$ and $-\bW$ have the same probability.
We refer to the latter property as \emph{blinding} (the identity of treatment).
Specifically,
$$\ts\sigma\in\Des=\braces{\R{\W}_+:\sum_{\bW\in\W}\sigma(\bW)=1,\,\sigma(\bW)=\sigma(-\bW)~\forall \bW\in\W},~ \W=\{\bW\in\{-1,1\}^n:\ip{\bW}{\bm 1}=0\}.$$
Given $\bX$ and given a design, $(\bY(-1),\bY(1),\bW,\bY^\obs)$ have a joint distribution (conditional on $\bX$).

Denoting by $\sigma$ the design we choose for every $\bX$ (hence a random variable), by repeating Theorem 7 of \citet{kallus2018optimal} we can shows that, since $\Eb{\hat\tau\mid \bX,\bY(-1),\bY(1)}=\tau$ due to blinding, $\hat\tau-\tau=\frac1n\ip{\bm W}{\bY(-1)+\bY(1)}$ by algebra, and $\Eb{\epsilon_i\mid \bX}=\Eb{\epsilon_i\epsilon_j\mid \bX}=0$ for $i\neq j$ by independence, we have
\begin{align}
\notag\varb{\hat\tau}&\ts
=\Efb{(\hat\tau-\tau)^2}+\varb{\tau}
=\frac1{n^2}\Efb{\ip\bW\bmu^2}+\frac1{n^2}\Efb{\ip\bW\bE^2}+
\frac2{n^2}\Efb{\ip\bW\bmu\ip\bW\bE}
+\varb{\tau}
\\&\ts
\label{eq:variancedecomp}
=\frac1{n^2}\Eb{B(\sigma,\bmu)}+\frac1{n^2}\sumi\varb{\epsilon_i}+\varb{\tau},\\
\notag\text{where}&\ts\quad B(\sigma,\bmu_0)=\sum_{\bW\in\W}\sigma(W)\ip\bW{\bmu_0}^2,
\end{align}
where we use the subscript $\vphantom{\sigma}_0$ to denote a dummy variable.
Notice that only the first term depends on the design, $\sigma$, and that $B(\sigma,\bmu)$ is measurable with respect to $\bX$ \emph{alone}. We of course do not know $\bmu$ so we consider a minimax framework. Given $\bX$ and some set $\M\subseteq\R n$ of potential values for $\bmu$, we define
$$\ts
\B(\sigma,\M)=\sup_{\bmu_0\in\M}B(\sigma,\bmu_0).
$$
The \emph{minimax-optimal design} is defined as the one minimizing $\B(\sigma,\M)$ (given $\bX$). Calling this the minimax design is based on the fact that, per \cref{eq:variancedecomp}, if we are given a random set $\M$ measurable with respect to $\bX$ and we choose the minimax design for \emph{each} $\bX$ then this experimental procedure minimizes the maximum variance of the (unbiased) estimator $\hat\tau$ over all measurable choices of $\bmu\in\M$.
This optimal design is called the MSOD in \citet{kallus2018optimal} to emphasize that it is a mixed strategy in this zero-sum game, \ie, it \emph{randomizes} over unit partitions. The PSOD is defined by \citet{kallus2018optimal} as the design that only randomizes over the assignments $\bW_0\in\W$ that minimize $\B(\bW_0\bW_0^\top,\M)$. Since this may not be minimax-optimal, it is given purely as a \emph{heuristic approximation} for the minimax-optimal MSOD and for the purpose of showing that various existing methods such as blocking, group mean matching, and nonbipartite pair matching are recovered as the PSOD for certain choices of $\M$.

Notice that since $\ip{\bW}{\bm1}=0$, we have $B(\sigma,\bmu_0)=B(\sigma,\bmu_0+\lambda\bm1)$ and so without loss of generality, it suffices to restrict $\M\subseteq\{\bmu_0:\ip{\bmu_0}{\bm1}=0\}$.
Next, notice that we can simplify $\B(\sigma,\M)=\sup_{\bmu_0\in\M}\ip{\bmu_0}{Q(\sigma)\bmu_0}$, where $Q(\sigma)=\sum_{\bW\in\W}\sigma(\bW)\bW\bW^\top$, \ie, $Q_{ij}(\sigma)=2\sigma(\{\bW:W_i=W_j\})-1$. Note $Q(\sigma)\bm1=\bm 0$.
In the following, we will often consider $\M=\M_K=\{K\bv:\ip{\bv}{K\bv}\leq \bd,\ip{K\bv}{\bm1}=0\}$ for a positive semidefinite $K=K^\top\in\R{n\times n}$,
for which $\B(\sigma,\M_K)=\bd\lambda_{\max}(K^{1/2}Q(\sigma)K^{1/2})$.

\section{The Optimality of Complete Randomization}

A natural next question is, what \emph{is} the minimax-optimal design? That, of course, depends on $\M$.
If we have no particular knowledge about $\bmu$, we should not constrain $\M$ in any informative fashion. However, $B(\sigma,\bmu_0)$ scales linearly with $\bmu_0$ so we must restrict it somehow else $\B(\sigma,\M)=\infty$ (equivalently, we must measure $B(\sigma,\bmu_0)$ relative to the \emph{magnitude} of $\bmu_0$). An uninformative restriction must be permutation symmetric, \ie, $\M$ is invariant to permutations of the coordinates of $\R n$.
An important permutation symmetric example is $\M_{\text{CR}}=\{\bmu_0:B(\sigma_\text{CR},\bmu_0)\leq\bd,\ip{\bmu_0}{\bm1}=0\}$,
where $\sigma_\text{CR}$ denotes the complete randomization design.
$\M_{\text{CR}}$ is of interest as it amounts to measuring one's variance relative to complete randomization's. A basic computation shows $B(\sigma_\text{CR},\bmu_0)=\ip{\bmu_0}{A_\text{CR}\bmu_0}$, 
where $A_\text{CR}=\frac{n}{n-1}(I-\frac{E}{n})$, $I$ is the identity matrix, and $E$ the matrix of ones, and that $\M_{\text{CR}}=\M_{K_{\text{CR}}}$ for $K_\text{CR}=\frac{n-1}{n}I$.
Other examples include $\M=\{\bmu_0:\fmagd{\bmu_0}_p\leq \bd\}$ for any $p$-norm.
Then, a basic exercise in convexity and symmetry shows that, whenever $\M$ is permutation symmetric, then the minimax-optimal design \emph{is} complete randomization. This is the ``no free lunch'' theorem of \citet[Theorem 1]{kallus2018optimal}: one cannot improve on complete randomization unless one is willing to assume \emph{structure}, \ie, some deviation from permutation symmetry. (While Theorem 1 of \citealp{kallus2018optimal} considers worst-case values of $\bY(0),\bY(1)$, here I take a more proper minimax approach, considering worst-case values of the \emph{parameters} $\bmu$, conditioned on $\bX$. I also do not assume identical unit distributions. The proof argument is exactly the same.)

\section{The Suboptimality of a Single Assignment}

Let us now consider a design that only uses a single partition of units and simply randomizes the identity of treatment, \ie, $\sigma_\text{single}(\bW_0)=\sigma_\text{single}(-\bW_0)=\frac12$ for some $\bW_0\in\W$. Then, $Q(\sigma_\text{single})=\bW_0\bW_0^\top$ and we can compute
$\B(\sigma_\text{single},\M_\text{CR})=\ip{\bW_0}{K_\text{CR}\bW_0}=\bd(n-1)$. In comparison, 
$\B(\sigma_\text{CR},\M_\text{CR})=\bd$ by construction. This says that, given $\bX$, for \emph{any} single $\bW_0$, there \emph{always} exists $\bmu_0\in\M_\text{CR}$ such that $B(\sigma_\text{single},\bmu_0)=\bd(n-1)$ and $B(\sigma_\text{CR},\bmu_0)=\bd$ (since $\M_\text{CR}$ is closed). In particular, there \emph{always} exists some $\bmu_0$ such that $B(\sigma_\text{single},\bmu_0)/B(\sigma_\text{CR},\bmu_0)=n-1$.
Now, take $\bd=\Theta(n)$, \ie, of order $n$. If $\bmu=\bmu_0$ and if $\varb{\epsilon_i},\varb{Y_i(1)-Y_i(0)}$ are bounded over $i=1,2,\dots$ (\eg, constant), then by \cref{eq:variancedecomp}, $\varb{\hat\tau_\text{CR}}=\Theta(1/n)$ while $\varb{\hat\tau_\text{single}}=\Theta(1)$. That is, CR has variance vanishing as $1/n$ and a design using a single partition has \emph{non-vanishing} variance. The existence of such $\bmu$ is a mathematical fact.

Example 1 of \citet{kallus2018optimal} provides an explicit construction of such $\bmu$ purely for illustration (\ie, not as a proof; the existence is already proven by computing $\B(\sigma_\text{single},\M_\text{CR})$). This particular example has $B(\sigma_\text{CR},\bmu)=\frac{4n^2}{n-1}=\Theta(n)$. Taking $\varb{\epsilon_i}=V_\epsilon$ and $\varb{Y_i(1)-Y_i(0)}=V_Y$ constant,
we have $\varb{\hat\tau_\text{CR}}=4/(n-1)+(V_\epsilon+V_Y)/n$ while $\varb{\hat\tau_\text{single}}=4+(V_\epsilon+V_Y)/n$. The example is specially constructed so that $\bW_0$ and $-\bW_0$ \emph{uniquely} optimize any scaled Euclidean distance between group means in $\bX$, \ie, $D_{\cv}(\bW)=\ip{\bW}{\bX \cv \bX^\top \bW}$ where $\bX\in\R{n\times d}$ and $\cv=\cv^\top\in\R{d\times d}$ is positive definite. \Eg, if $\cv$ is the inverse sample covariance matrix, this gives the Mahalanobis distance. (The example is also appealing as it recovers the worst-case behavior of nonbipartite pair matching and blocking. Note also that in Journal production, a typo in Example 1 was introduced overlooked in the proofing where some $t$'s are typeset as $\iota$ and two $b$'s were dropped; the typo does not appear in an earlier arXiv preprint version. The correct construction is $b\in\mathbb N$, $n=2^b$, $X_i=\sum_{t=0}^{b-\max\{2,\log_2(i)\}}(-1)^{\lceil i/2^{b-t-1}\rceil}\times2^{
-2^{b-1}+2^{b-t-1}+(i-1~\op{mod}~2^{b-t-1})
}$.)

\citet{johanssoncomment} cite this example and \emph{incorrectly} claim that $\bW_0$ and $-\bW_0$ do \emph{not} uniquely optimize the Mahalanobis distance. In fact, they misquote the example. While they incorrectly claim that the construction has $\bX=(2^0,\dots,2^{n/2-1},-2^0,\dots,-2^{n/2-1})$, Example 1 of \citet{kallus2018optimal} clearly provides a \emph{different}, much more involved formula for $X_i$ and writes that ``This rather complicated construction essentially yields $\bX\approx\op{round}(\bX)=(2^0,\dots,2^{n/2-1},-2^0,\dots,-2^{n/2-1})$ with just enough perturbation so that the assignment [$\bW_0=(-1,1,\dots,-1,1)$] uniquely minimizes Mahalanobis distance between group means'' (where the fact that $-\bW_0$ is also optimal is implicit since we always blind the identity of treatment so we only discuss the unit partitions). Of course, if we had considered $\op{round}(\bX)$ as our pre-treatment variables, this would not be the uniquely optimal partition, but these are \emph{not} the pre-treatment variables in Example 1 of \citet{kallus2018optimal}.
More generally, it is a fact, per the above, that whenever $\bW_0,-\bW_0$ uniquely optimize the Mahalanobis distance, there will \emph{always} exist some mean-outcome vector $\bmu$ such that the design $\sigma_\text{Maha-opt}$ randomizing over all optimizers of Mahalanobis distance will have $B(\sigma_\text{Maha-opt},\bmu)/B(\sigma_\text{CR},\bmu)=n-1$. Example 1 of \citet{kallus2018optimal} is just one (correct) explicit example. The claim of \citet{johanssoncomment} that ``The mistake of \citet{kallus2018optimal} stems from the incorrect assumption that the allocation [$\bW_0=(-1,1,\dots,-1,1)$] uniquely minimizes the Mahalanobis distance for all $n$'' is patently false: they consider a \emph{different} set of covariates $\bX$ than \citet{kallus2018optimal}.

\section{The Optimality of Restricted Randomization: The Mixed-Strategy Optimal Design}\label{sec:optrestricted}

The next natural question is, when is something \emph{different} from CR minimax-optimal? That, again, depends on $\M$.
Consider the case of $\M=\M_K$. Then the optimal design is that which minimizes $\B(\sigma,\M_K)=\bd\lambda_{\max}(K^{1/2}Q(\sigma)K^{1/2})$, so it depends on the spectrum of $K$. In one extreme, $\M_K$ is permutation symmetric, in which case the whole spectrum of $K$ must be concentrated in a single value (aside from the eigenvector $\bm 1$), $\M_K=\M_{\lambda I}$ for some $\lambda$, and CR becomes optimal. In the other extreme, $K=\bv\bv^\top$ is of rank one, in which case $\lambda_{\max}(K^{1/2}Q(\sigma)K^{1/2})=\ip{\bv}{Q(\sigma)\bv}$, and the \emph{single} partition $\bW_0,-\bW_0$ that solves $\min_{\bW\in\W}\abs{\ip{\bW}{\bv}}$ becomes the optimal design. In between these extremes, when $K$ has a dispersed spectrum, something between perfectly partitioning a single vector and complete randomization is optimal: the MSOD of \citet{kallus2018optimal}.

To motivate other constructions of $\M$ suppose that there exists $f:\X\to\Rl$ such that $\mu_i=f(X_i)$. This, for example, would be guaranteed if units were identically distributed. Now let $\M_\F=\{(f(X_1),\dots,f(X_n)):f\in\F\}$ where $\F\subseteq[\X\to\Rl]$ is some class of functions. Suppose that $\F$ is the unit ball of a reproducing kernel Hilbert space (RKHS), \ie, for a positive semidefinite kernel $\K:\X\times\X\to\Rl$, $$\ts\F=\F_{\K}=\{x\mapsto\sum_{i=1}^\infty \gamma_i\K(z_i,x):\bm z\in\X^{\mathbb N},\bm\gamma\in\R{\mathbb N},\sum_{i=1}^\infty\gamma_i^2\K(z_i,z_i)<\infty,\sum_{i,j=1}^\infty\gamma_i\gamma_j\K(z_i,z_j)\leq1\}.$$
Then one can show that $\M_{\F}=\M_K$, where $K_{ij}=\K(X_i,X_j)$. Examples of positive semidefinite kernels when $\X=\R d$, as given in \citet{kallus2018optimal}, are linear $\K(x,x')=\ip{x}{\cv x'}$, polynomial $\K(x,x')=(1+\ip{x}{\cv x'})^m$, and Gaussian $\K(x,x')=\exp(-\ip{x-x'}{\cv(x-x')})$, all for some positive semidefinite $\cv=\cv^\top\in\R {d\times d}$. This offers the researcher a flexible modeling framework and clearly connects assumptions on the structure of the CEF $f$ to optimal design. The Gaussian kernel is notable for being a \emph{universal} kernel: the span of $\F$ is dense in continuous functions in $L_\infty$ (or in all functions in $L_p$). This ensures model-free consistency even without assuming $f\in\op{span}(\F)$ \citep[Theorem 13]{kallus2018optimal}.

Consider next the linear kernel with a positive definite $\cv$. We can then rewrite $\F_{\K}=\{x\mapsto \ip\beta x:\ip{\beta}{\cv^{-1}\beta}\leq1\}$, \ie, the set of CEFs are the linear functions with coefficients bounded in $\cv$-scaled norm. Notice that in this case, $K=\bX \cv\bX^\top$. Now, if $\sigma_\text{single}$ has the single partition $\bW_0,-\bW_0$ then $\B(\sigma_{\text{single}},\M_K)=\bd\cdot D_{\cv}(\bW_0)$, \ie, the $\cv$-scaled Euclidean distance between group means. Therefore, among \emph{single}-partition designs, the optimal one, \ie, the PSOD, minimizes the distance in experimental group means in pre-treatment variables; Mahalanobis distance if $\cv$ is the inverse sample covariance. However, if $d\geq2$ then $K$ is generally of rank higher than one and a single partition is \emph{not} minimax-optimal. This means that, unlike the characterization of \citet{johanssoncomment}, even in the simple linear-CEF setting that recovers Mahalanobis mean matching, a single partition is \emph{not} optimal and randomization beyond just blinding \emph{is necessary} to achieve minimax-optimality, that is, the so-called MSOD proposed in \citet{kallus2018optimal}. So even when we care about balancing experimental group means, and even when we are optimizing only for minimax variance, we should \emph{still} randomize over partitions.

There is also an easy way to trade-off optimality for a given single CEF and additional randomization. Suppose we have a guess $f_0$ for the CEF. If we set $\M=\{\bmu_0\}$ to a singleton of $\bmu_0=(f_0(X_1),\dots,f_0(X_n)$, then $\B(\sigma,\M)=\B(\sigma,\M_{K_0})$ where $K_0=\bmu_0\bmu_0^\top$ and the optimal design is the single partition that minimizes the subset sum differences for the vector $\bmu_0$, as in the rank-one example in the first paragraph of this section.
This is optimal if $f=f_0$, but if we knew the outcome CEFs we would not need an experiment to begin with.
If we want to introduce additional randomization (\eg, if we are unsure of our guess $f_0$), we may use $\M_{K_\lambda}$ for $K_\lambda=K_0+\lambda I$, \ie, wash out the spectrum of $K_0$. While $\lambda=0$ recovers the perfect partitioning of $\bmu_0$, $\lambda\to\infty$ recovers complete randomization. This exactly corresponds to considering the CEF set given by $\F=\{f_0\}+\lambda\F_{\K_\delta}$ for the Dirac Kernel $\K_\delta(x,x')=\indic{x=x'}$ (note this is not a Mercer kernel).
More generally, given any $K$, we can regularize the corresponding minimax-optimal design toward more randomization by using $K+\lambda I$ instead.
Alternatively, given any $\F_0$ (\eg, an RKHS ball, $\F_\K$) one can also use $\M_\F$ for $\F=\F_0+\lambda\F_{\K_\delta}$ or $\F=\F_0+\{f:\magd f_\infty\leq \lambda\}$. Expanding $\F_0$ by arbitrary bounded functions washes out more structure as we increase $\lambda$.
The result is similar to \citet{kapelner2020harmonizing} but using a proper minimax framework on parameters rather than introducing adversarial choice of random variables.

Given $K$, the minimax-optimal design (the \emph{randomized} MSOD) solves $\min_{\sigma\in\Des}\lambda_{\max}(K^{1/2}Q(\sigma)K^{1/2})=\min_{Q\in\Q}\lambda_{\max}(K^{1/2}QK^{1/2})$, where $\Q=\{Q(\sigma):\sigma\in\Des\}$. This, however, may be a difficult optimization problem. For that reason, \citet{kallus2018optimal} provides an outer approximation $\Q\subseteq\{Q~\text{positive semidefinite}:\op{diag}(Q)=\bm 1,Q\bm 1=\bm 0\}$, as well as an inner approximation $\Q\supseteq\{Q(\sigma):\sigma\in\Des,\sum_{\bW\in\mathcal W_0}\sigma(\bW)=1\}$, where $\mathcal W_0=-\mathcal W_0\subset\W$ is a given set. Both approximations give rise to tractable semidefinite programs.

\section{When Is a Single Partition Optimal?}\label{sec:whensingle}

\Cref{sec:optrestricted} shows that a single partition (randomizing between $\bW_0,-\bW_0$) is minimax-optimal for $\M_K$ when $K$ is rank one. Otherwise, it is generally \emph{suboptimal}.
It is worth mentioning that in the minimax framework it is optimal to randomize only because the researcher plays first and Nature second in this zero-sum game, in the sense that the researcher first announces her mixed (\ie, randomized) strategy (distribution over $\bW$) but not the specific $\bW$ she will play, and then Nature choose an action adversarially to maximize the expected loss averaged over $\bW$'s from the researcher's strategy. Once the first player announces her strategy, the second player does not benefit from randomization. Indeed, if the roles were reversed, and first Nature announced its distribution over $\bmu$ and then the researcher chose a design, then the researcher would not benefit from randomization (for minimizing squared error). This is \emph{precisely} the Bayesian setting: the distribution announced by Nature is the prior over $\bmu$. In this setting, the researcher need not randomize to minimize squared error. But this assumes we know a prior and is therefore unappealing in a controlled experiment, where we can potentially have assumption-free correct causal inference if we randomized. The minimax framework may therefore be preferable.

\section{Optimizing for Randomization Inference}

The minimax-optimality framework deals with optimizing \emph{precision}, but does not explicitly handle inference. One can attempt to study the marginal sampling distribution of $\hat\tau$ but that may prove difficult. A more convenient and assumption-free approach may be to use a randomization (aka Fisher exact) test. I here provide an extension of the MSOD that constrain the optimization to ensure designs that support randomization testing at a given significance.

A randomization test can be run for \emph{any} design to test the sharp null hypothesis, $H_0:\bY(-1)=\bY(1)$. First fix some test statistic $s(\bW,\bY^\obs)$ (it can also depend on $\bX$ since everything is conditional on $\bX$).
\Eg, the absolute mean difference $s(\bW,\bY^\obs)=\abs{\hat\tau}$, or the absolute value of the two-sample $t$-statistic (either pooled variance or Welch's).
Then, after assignment and treatment, we record $s(\bW,\bY^\obs)$. Under $H_0$, $\bY^\obs$ would also be what we observed if we made another treatment assignment, so the distribution of $s(\bW,\bY^\obs)$ given $\bX$ is given by $s(\bW_0,\bY^\obs)$ where $\bW_0\sim\sigma$. This gives the $p$-value $p=\sum_{\bW_0\in\W}\sigma(\bW_0)\indic{s(\bW_0,\bY^\obs)\geq s(\bW,\bY^\obs)}$, which can be approximated by Monte-Carlo simulation from $\sigma$. By construction, if we only reject $H_0$ when $p\leq\alpha$ then our type-I error rate must be at most $\alpha$.

A concern, however, is the power of the test. One may hope that if precision is high, then power would also be high. But, if we only randomize over a single partition, then for the above examples of two-tailed statistics, we will always have $p=1$ and we never reject the null.
We must therefore ensure that each assignment only occurs with probability at most $\alpha/2$ (focusing on two-sided statistics).
The MSOD, despite being randomized, may or may not have this property for a given $\alpha$. 

I therefore propose the \emph{inference-constrained} MSOD, which,
given $K$ and $\alpha$, solves:
\begin{equation}\label{eq:constrainedMSOD}\ts
\min_{\sigma\in\Des\;:\;\sigma(W)\leq \frac\alpha2\,\forall W\in\W}~\lambda_{\max}\prns{\sum_{\bW\in\W}\sigma(\bW)K^{1/2}\bW\bW^\top K^{1/2}}.
\end{equation}
The constraint $\sigma(W)\leq \frac\alpha2$ ensures that each assignment has probability at most $\alpha/2$ so that, if chosen, the randomization test can potentially return $p\leq\alpha$ if the statistic is indeed extreme. \Cref{eq:constrainedMSOD} can be a difficult optimization problem so I propose the following approximation based on \citet[Algorithm 3]{kallus2018optimal}. Set $\W_1=\W\cap\{\bW:W_1=1\}$; for $t=1,\dots,T$, solve $\bW_t\in\argmin_{\bW\in\W_t}\bW^\top K\bW$ and set $\W_{t+1}=\W_t\cap\{\bW:\bW_t^\top\bW\leq n-4\}$. Then $\bW_1,-\bW_1,\dots,\bW_T,-\bW_T$ are the top $2T$ solutions to $\min_{\bW\in\W}\bW^\top K\bW$ (the top two, if unique, gives the PSOD). Each optimization problem is a binary optimization problem with a convex-quadratic objective and linear constraints and can be solved with off-the-shelf solvers such as Gurobi. Then 
let $U\in\R{n\times T}$ have the columns $\bW_t$ and solve
\begin{equation}\label{eq:constrainedMSOD2}\ts
\min_{\lambda\in\Rl,\,\bv\in\R T_+\;:\;\bv\leq \alpha\bm1,\,\ip{\bv}{\bm1}=1,\,\lambda I-K^{1/2}U \op{diag}(\bv) U^\top K^{1/2}~\text{is positive semidefinite}}~\lambda,
\end{equation}
and set $\sigma(\bW_t)=\sigma(-\bW_t)=v_t/2$. \Cref{eq:constrainedMSOD2} is a tractable semidefinite program.
Notice we need at least $T\geq1/\alpha$ for \cref{eq:constrainedMSOD2} to be feasible. If $1/\alpha$ is integral, setting $T=1/\alpha$ forces \cref{eq:constrainedMSOD2} to choose the design that uniformly randomizes over the top $2T$ solutions to $\min_{\bW\in\W}\bW^\top K\bW$. This latter alternative approach was considered in \citet[Example 4]{kallus2018optimal} but was found empirically less powerful than a bootstrap test. Focusing solely on randomization tests, \cref{eq:constrainedMSOD} is exactly the minimax-optimal design for optimizing variance subject to the constraint of no single assignment occurring more often than $\alpha/2$. For larger $T$, \cref{eq:constrainedMSOD2} provides a good approximation of this design.

\section{The Suboptimality of Rerandomization}

\citet{morgan2012rerandomization} proposed the design that uniformly randomizes over $\{\bW\in\W:D_{\cv}(\bW)\leq a\}$, which they operationalize by repeatedly sampling $\bW$ uniformly from $\W$ until $D_{\cv}(\bW)\leq a$, termed rerandomization. Specifically, they use $\cv=\op{cov}(\bX)^{-1}$ (so $D_{\cv}$ is Mahalanobis distance) and recommend setting $a=F^{-1}_{\chi_d^2}(p_a)$, where $F_{\chi_d^2}$ is the cumulative distribution function of the $\chi^2$-distributions with $d$ degrees of freedom (assuming $\bX\in\R {n\times d}$) and $p_a\in(0,1)$ is a target acceptance probability, which this procedure approximates. This is particularly notable for nicely formalizing and theoretically characterizing what was previously a haphazard practice of researchers repeatedly clicking ``Recalculate'' (F9) in Excel whenever the unconstrained randomization of units obtained appeared ``imbalanced'' or ``undesirable'' (this is what ``historically haphazard'' refers to in \citealp{kallus2018optimal}; not to the method of \citealp{morgan2012rerandomization} nor the use of Mahalanobis or linear projections, as suggested by \citealp{johanssoncomment}). 
\citet{johanssoncomment} highlight that rerandomization is also notable for both improving precision \emph{and} enabling randomization inference, as long as $a$ is chosen so that $\abs{\{\bW\in\W:D_{\cv}(\bW)\leq a\}}\geq2/\alpha$.

It is important to note, however, that in our minimax framework the rerandomization design is \emph{not} minimax-optimal for $\M=\M_{\bX\cv\bX^\top}$. The only exception is the case $d=1$, where a single partition (\ie, the PSOD) is optimal as discussed in \cref{sec:whensingle}; this is equivalent to setting $a=\min_{\bW\in\W}D_{\cv}(\bW)$ in the above (which \citealp{johanssoncomment} also refer to as ``optimal rerandomization,'' perhaps oxymoronically). However, in practice, we generally have $d\geq2$, in which case, firstly, the minimax-optimal design (the MSOD) requires \emph{more} randomization than a single partition, and the rerandomization design is \emph{not} minimax-optimal for \emph{any} value of $a$. In particular, even if set $a$ such that $\abs{\{\bW\in\W:D_{\cv}(\bW)\leq a\}}=2/\alpha$, not only do we still not obtain the optimal design, but using rejection sampling will also have very bad running time as it is solving constraint satisfaction by brute force Monte Carlo. In particular, rerandomization requires roughly $1/p_a=1/F_{\chi_d^2}(a)$ samples. This is not an issue if $p_a$ is fixed with $n$, \eg, $0.1$, but it also means that we are uniformly randomizing over very many partitions so we are even \emph{less} optimal. If $p_a$ is decreasing with $n$ so that $\abs{\{\bW\in\W:D_{\cv}(\bW)\leq a\}}$ is roughly constant, this means we need an exponentially growing number of samples. This is even worse at inference time when we need to again sample multiple times from this design, each time needing exponentially many samples.

\section{Concluding Remarks}

I here studied the minimax-optimal design when conditional mean outcomes may vary in a given set. This more clearly and formally positions the framework of \citet{kallus2018optimal} as a minimax one and makes clear that the MSOD defined therein is the minimax-optimal design. This also demonstrated that the design that is minimax-optimal for variance \emph{does} still randomize over more than one partition. Since this, nonetheless, only optimizes for precision, it may not ensure sufficient uniformity to ensure randomization inference at any given significance $\alpha$ has power to detect violations of null. I therefore proposed the new inference-constrained MSOD (\cref{eq:constrainedMSOD}) and a tractable heuristic for it (\cref{eq:constrainedMSOD2}). While rerandomization designs with sufficiently large $a$ enable randomization inference, they do not optimize any principled error objective. Instead, the inference-constrained MSOD is minimax-optimal for precision among all designs with sufficient uniformity to enable randomization inference at a given significance $\alpha$.

I thank and congratulate \citet{johanssoncomment} for a thought-provoking paper and for highlighting the importance of randomization inference. However, I find it made a few errors, which I here used the opportunity to set straight: that minimax variance optimality does not mean using a single unit partition, \ie, the minimax-optimal design randomizes beyond blinding and that this was proposed in \citet{kallus2018optimal}; that one can enable randomization inference while still targeting a principled objective; and that Theorem 1 and Example 1 of \citet{kallus2018optimal} are correct.
Their work nonetheless inspired my proposal herein of inference-constrained MSODs and I welcome the continued vigorous conversation.

\bibliographystyle{chicago}
\bibliography{OptimalityOfRandomization}

\end{document}